\documentclass{jpp}
\pdfoutput=1
\usepackage{setspace}
\usepackage{graphicx}
\usepackage{soul,xcolor}
\setstcolor{red}
\usepackage{graphicx}
\graphicspath{ {./Figures/} }
\usepackage[utf8x]{inputenc}
\usepackage[T1]{fontenc}
\usepackage{amsmath}
\usepackage{appendix}
\usepackage{subcaption}    
\usepackage{float}
\usepackage[normalem]{ulem}
\DeclareMathOperator{\sech}{sech}
\shorttitle{The On-Axis Magnetic Well and Mercier's Criterion}
\shortauthor{P. Kim, R. Jorge, W. Dorland}
\usepackage[capitalize]{cleveref}

\title{The On-Axis Magnetic Well and Mercier's Criterion for Arbitrary Stellarator Geometries}

\author{P. Kim\aff{1}
  \corresp{\email{pkim1236@umd.edu}},
  R. Jorge\aff{1}
 \and W. Dorland\aff{1,2}}

\affiliation{\aff{1}Institute for Research in Electronics and Applied Physics, University of Maryland, College Park, MD 20742, USA
\aff{2}Department of Physics, University of Maryland, College Park, MD 20742, USA}

\begin{document}

\maketitle

\begin{abstract}
A simplified analytical form of the on-axis magnetic well and Mercier's criterion for interchange instabilities for arbitrary three-dimensional magnetic field geometries is derived. For this purpose, a near-axis expansion based on a direct coordinate approach is used by expressing the toroidal magnetic flux in terms of powers of the radial distance to the magnetic axis. For the first time, the magnetic well and Mercier's criterion are then written as a one-dimensional integral with respect to the axis arclength. When compared with the original work of Mercier, the derivation here is presented using modern notation and in a more streamlined manner that highlights essential steps. Finally, these expressions are verified numerically using several quasisymmetric and non-quasisymmetric stellarator configurations including Wendelstein 7-X.
\end{abstract}


\section{Introduction}
In order to operate a functional nuclear fusion reactor, the high-temperature plasma must be confined in a stable equilibrium against various fluctuations over extended periods of time. Small perturbations in the plasma can lead to instabilities that ultimately can cause the release of large bursts of energy and heat to the vessel's wall \citep{Freidberg2007PlasmaEnergy}. In tokamaks, the plasma is confined via a strong magnetic field generated using axisymmetric coils and an externally driven longitudinal current. This makes steady-state operation in tokamaks challenging to achieve. Stellarator devices, on the other hand, circumvent this problem by using asymmetric coils carefully designed to produce an equilibrium magnetic field without the need for an additional solenoid or plasma current. Such magnetic fields are usually obtained using reduced plasma models such as the ideal magnetohydrodynamic (MHD) equilibrium equation \citep{Freidberg2014IdealMHD}%
\begin{equation}\label{ideal MHD}
    \mathbf{J} \times \mathbf{B} = \nabla p,
\end{equation}
where $\mathbf{B}$ is the magnetic field vector, $\mathbf{J} = \nabla \times \mathbf{B}$ is the plasma current, and $p$ is the plasma pressure.
Due to the breaking of axisymmetry, the number of degrees of freedom needed to describe the equilibrium magnetic field in a stellarator increase by approximately an order of magnitude compared to those of tokamaks \citep{Boozer2015StellaratorDesign}. This has paved the way for stellarator optimization techniques to obtain good plasma confinement properties such as low neoclassical transport and MHD stability \citep{Helander2014TheoryFields}. For this purpose, modern optimized stellarators such as W7-X \citep{Geiger2015PhysicsW7-X} and HSX \citep{Anderson1995TheStatus} were designed using computationally expensive numerical optimization algorithms that are highly dependent on the initial point of the parameter space and the particular set of weights used for the objective functions. Due to the high-dimensionality of the available parameter space, it is often too numerically intensive to ensure that a global minimum of the objective function was found \citep{Boozer2015StellaratorDesign}. Furthermore, it is easier to gain physical insight from analytical formulations, such as the near-axis one used here, than from numerical tools.

One of the most commonly employed conditions for MHD stability used in stellarator optimization studies is Mercier's criterion \citep{Mercier1964EquilibriumAxis, Beidler1990, Anderson1995TheStatus, Drevlak2019OptimisationROSE}. This criterion is often simplified using the fact that in the limit of large aspect ratio and small pressure $p$ compared to the magnetic pressure $B^2/2$, i.e., $\beta = 2p/B^2 \ll 1$, the dominant term in the criterion is given by $p'(\psi)V''(\psi)$, where $V$ is the volume enclosed by a toroidal magnetic flux surface and $\psi$ is the toroidal magnetic flux. When $p'(\psi) < 0$ and $V''(\psi) < 0$, the equilibrium is said to possess a magnetic well, and is approximately stable against interchange-like perturbations. For this reason, the magnetic well is often used as the main proxy for ideal MHD stability \citep{Beidler1990, Anderson1995TheStatus, Drevlak2019OptimisationROSE}. Additionally, the lowest order Mercier's criterion in the distance to the axis for ideal modes coincides with the lowest order Glasser, Greene, and Johnson's stability criterion for resistive modes \citep{glasser1975, Landreman2020MagneticAxis}. 

In this work, we derive an analytical expression for Mercier's criterion (and hence the magnetic well) on the magnetic axis of a stellarator. One of the main advantages to using an analytical expression for optimization quantities near the magnetic axis is that such expressions can quickly produce several sets of reasonable initial parameters that may allow expensive optimization codes to avoid the same local minimum. As shown in \citet{Landreman2019OptimizedConstruction}, an expansion near the magnetic axis can accurately describe the inner core of optimized experimental devices using only a reduced number of free parameters.
In the near-axis formulation, an expansion in terms of powers of the inverse aspect ratio is employed
\begin{equation}
    \epsilon = \frac{a}{R} \ll 1,
\end{equation}{}%
where $a$ is the maximum perpendicular distance from the magnetic axis to a particular flux surface and $1/R$ is the minimum curvature of the magnetic axis. In this work, we follow the direct coordinate approach pioneered by \citet{Mercier1964EquilibriumAxis} and \citet{Solovev1970PlasmaSystems}, and extended to arbitrary order in \citet{Jorge2020Near-axisAxis}. In the direct approach, $\psi$ is expressed in terms of the Mercier coordinates $(\rho,\theta,s)$, with $\theta$ being the polar angle in the plane normal to the axis and $s$ the arc length of the axis. We note that in comparison to \citet{Jorge2020Near-axisAxis}, there is no subsidiary expansion for small $\beta$ and the calculations are performed at arbitrary values of $\beta$. This formulation is applied to Mercier's criterion to find its lowest order component. The near-axis expansions have already been used in the past to study MHD stability. While the first derivation of Mercier's criterion is presented in \citet{Mercier1964EquilibriumAxis}, in \citet{Solovev1970PlasmaSystems} the specific volume $V'(\psi)$ is calculated in vacuum, while its non-vacuum counterpart is derived in \citet{Lortz1977}. Most recently, in \citet{Landreman2020MagneticAxis}, Mercier's criterion is derived using the inverse-coordinate approach in Garren-Boozer coordinates where the spatial position vector $\mathbf{r}$ is expressed in terms of magnetic coordinates including $\psi$. While \citet{Landreman2020MagneticAxis} derives a formula for arbitrary geometries, i.e., both quasisymmetric and non-quasisymmetric geometries, due to the relation between this approach and Boozer coordinates, a numerical method using this approach has only been developed for quasisymmetric stellarators. Therefore, when compared to \citet{Landreman2020MagneticAxis}, this work provides an independent derivation of the on-axis magnetic well and Mercier's criterion for both vacuum and non-vacuum cases, and verifies it against non-quasisymmetric devices. Furthermore, the designs used to calculate the magnetic wells in \citet{Mercier1964EquilibriumAxis}, \citet{Solovev1970PlasmaSystems}, and \citet{Lortz1977} are limited to simpler stellarator geometries, such as those with circular axis shapes or circular surface cross-sections. Such works also employ older notation and were not the subject of independent verification efforts. In this work, we intend to put the expressions for the on-axis magnetic well and Mercier's criterion in firmer footing using modern notation and employing a more streamlined approach. Finally, we perform the first verification of Mercier's formulation for arbitrary stellarator geometries by evaluating the magnetic well for W7-X and two quasisymmetric stellarator designs from \citet{Landreman2019ConstructingOrder}.

This paper is organized as follows. In Section \ref{sec:Near-axis}, we describe the direct coordinate near-axis framework to express the toroidal flux, as well as the current density and magnetic field in terms of geometric quantities related to the magnetic axis and toroidal flux surface. Then, in Section \ref{sec:magnetic well}, we derive the on-axis magnetic well. The lowest order component of Mercier's criterion is derived Section \ref{sec:mercier's criterion}, while in Section \ref{sec:circular cross section} the expression for the magnetic well is compared to its inverse approach counterpart for the case of a tokamak with circular cross sections. Finally, in Section \ref{sec:numerical verification}, we numerically verify our expression for the magnetic well and the lowest order Mercier's criterion. The conclusions follow.

\section{Near-Axis Expansion}\label{sec:Near-axis}
In this section, we introduce the near-axis expansion using the direct coordinate approach first derived in \citet{Mercier1964EquilibriumAxis}. We denote the magnetic axis as the curve $\mathbf{r}_0(s)$ having total length $L$, with $s$ the arclength. The tangent $\mathbf{t}(s) = \mathbf{r}_0'(s)$, the normal $\mathbf{n}(s) = \mathbf{t}'(s)/\kappa$ and the binormal $\mathbf{b}(s) = \mathbf{t} \times \mathbf{n}$ unit vectors form the orthogonal Frenet-Serret frame \citep{Spivak1999} which satisfy the first-order Frenet-Serret equations $\mathbf{t'}(s) = \kappa \mathbf{n}$,  $\mathbf{n'}(s) = -\kappa \mathbf{t} + \tau \mathbf{b}$, and $\mathbf{b'}(s) = -\tau \mathbf{n}$. The curvature $\kappa$ and the torsion $\tau$ are found using
\begin{equation}\label{curvature}
    \kappa(t) = \frac{\lvert \mathbf{r}_0'(t) \times \mathbf{r}_0''(t)\rvert}{\lvert \mathbf{r}_0'(t)\rvert^2},
\end{equation}
and
\begin{equation}\label{torsion}
    \tau(t) = \frac{\left[\mathbf{r}_0'(t) \times \mathbf{r}_0''(t)\right] \cdot \mathbf{r}_0'''(t)}{\lvert \mathbf{r}_0'(t) \times \mathbf{r}_0''(t)\rvert^2},
\end{equation}%
with $t$ any quantity that parameterizes $\mathbf{r}_0$. Finally, we note that the triad $(\mathbf{t},\mathbf{n},\mathbf{b})$ is uniquely determined by the Frenet-Serret equations above. 

In order to describe a point outside the magnetic axis, we consider a plane between an arbitrary point $\mathbf{r}$ and a point on the magnetic axis $\mathbf{r}_0$ such that the axis is normal to that plane. The position vector $\mathbf{r}$ can then be written as
\begin{equation}
    \mathbf{r} = \mathbf{r}_0(s) + \rho\cos\theta\mathbf{n}(s) + \rho\sin\theta\mathbf{b}(s),
\end{equation}%
with $\rho$ the distance between $\mathbf{r}$ and $\mathbf{r}_0$ and $\theta$ the angle between the normal and $\mathbf{r}-\mathbf{r}_0$. Alternatively, this plane can be visualized as a Cartesian plane mapped onto the $(\mathbf{n},\mathbf{b})$ plane with $x = \rho\cos\theta$ and $y = \rho\sin\theta$. We now introduce the angle $\omega = \theta + \gamma(s)$, where $\gamma$ is the integrated torsion $\gamma(s) = \int_0^s \tau(s') ds'$, which renders the coordinates $(\rho,w,s)$ orthogonal. The Jacobian in $(\rho,\omega,s)$ coordinates is then given by $\sqrt{g} = \rho h_s = \rho(1-\kappa\rho\cos({\omega - \gamma(s)})$.

We decompose the magnetic field as $\mathbf{B} = B_\rho\mathbf{e}_\rho + B_\omega\mathbf{e}_\omega + B_s\mathbf{e}_s$, with $\mathbf{e}_\rho = \cos \theta\mathbf{n} + \sin \theta\mathbf{b}$, $\mathbf{e}_\omega = -\sin \theta\mathbf{n} + \cos \theta\mathbf{b}$, and $\mathbf{e}_s = \mathbf{t}$, and similarly for the current density $\mathbf{J} = \nabla \times \mathbf{B}$. Each scalar function is then expanded in a power series in terms of $\epsilon\rho$, i.e.

\begin{equation}\label{B field expansion}
    B_\rho = \sum_{n=0}^{\infty} B_{\rho n}(\omega,s)\epsilon^n \rho^n.
\end{equation}
We note that in a vacuum, as the current density vanishes, $\mathbf{B}$ is a curl-free field, i.e., $\nabla \times \mathbf{B} = 0$. In this case, a scalar potential function $\phi$ with $\mathbf{B} = \nabla\phi$ satisfying Laplace's equation can be used. In orthogonal Mercier's coordinates $(\rho, \omega, s)$, Laplace's equation can be written as
\begin{equation}\label{laplace}
    \nabla^2\phi = \frac{1}{\epsilon^2\rho}\frac{\partial}{\partial\rho}\left(h_s\rho\frac{\partial \phi}{\partial\rho}\right) + \frac{1}{\epsilon^2\rho^2}\frac{\partial}{\partial\omega}\left(h_s\frac{\partial\phi}{\partial\omega}\right) + \frac{\partial}{\partial s}\left(\frac{1}{h_s}\frac{\partial\phi}{\partial s}\right) = 0.
\end{equation}%
A more detailed treatment of Eq. \eqref{laplace} can be found in \citet{Jorge2020Near-axisAxis}. 

In order to simplify the derivation and reduce the number of scalar functions to compute, we write the magnetic field using a Clebsch representation \citep{Helander2014TheoryFields}
\begin{equation}\label{clebsch}
    \mathbf{B} = \nabla\psi \times \nabla\alpha,
\end{equation}%
where $\alpha$ is a field line label and $\psi$ the magnetic toroidal flux 
\begin{equation}\label{eq:torflux}
    \psi = \frac{1}{L}\int(\mathbf{B} \cdot \nabla s)dV.
\end{equation}
Both $\psi$ and $\alpha$ are expanded in powers of $\epsilon\rho$ similarly to Eq. \eqref{B field expansion}. An equation for the toroidal flux $\psi$ can be found by writing the constraint $\mathbf{B} \cdot \nabla \psi = 0$ in Mercier's coordiantes, yielding
\begin{equation}\label{psi_diffeq}
    B_\rho\frac{\partial\psi}{\partial\rho} + B_\omega\frac{1}{\rho^2}\frac{\partial\psi}{\partial\omega} + B_s\frac{\epsilon}{h_s}\frac{\partial\psi}{\partial s} = 0.
\end{equation}
Assuming that $\psi$ is analytic close to the magnetic axis [see \citet{Kuo-Petravic1987NumericalHamiltonian} and \citet{Jorge2020Near-axisAxis} Eqs. 2.12-2.14 for more details on the Fourier series expansion of analytic functions near the axis in the Frenet-Serret frame], we write each component of $\psi$ as
\begin{equation}\label{psi analytic}
    \psi_n(\omega,s) = \sum_{p=0}^{n} \psi_{np}^{c}(\omega,s)\cos p\theta + \psi_{np}^{s}(\omega,s)\sin p\theta,
\end{equation}%
where the angle $\theta=\omega-\gamma(s)$ is used as all physical quantities must be periodic in $\theta$ (but not necessarily in $\omega$).
Noting that $\psi = O(\rho^2)$ from the constraint in Eq. \eqref{eq:torflux}, we write its lowest order components as
\begin{equation}\label{second_order}
    \psi_2 = \frac{B_0\pi}{\sqrt{1-\mu^2}}(1 + \mu\cos 2u),
\end{equation}%
and
\begin{equation}\label{third_order}
    \psi_3(\omega,s) = \psi_{31}^{c}\cos u + \psi_{31}^{s}\sin u + \psi_{33}^{c}\cos 3u + \psi_{33}^{s}\sin 3u,
\end{equation}%
where $B_0$ is the on-axis magnetic field, $u = \theta + \delta(s)$ and $\delta(s)$ is chosen such that there is no $\sin 2u$ term in Eq. \eqref{second_order}. In Eq. \eqref{second_order}, and $\mu$ represents the eccentricity \citep{Solovev1970PlasmaSystems}. We remark that the analyticity condition in Eq. \eqref{psi analytic} also applies for non-vacuum cases, so the forms of $\psi_2$ and $\psi_3$ are the same for finite $\beta$.

The divergence-free condition associated with the plasma current, $\nabla \cdot \mathbf{J} = 0$, allows us to write $\mathbf{J}$ using a Clebsch representation similar to Eq. \eqref{clebsch},
\begin{equation}\label{clebsch current}
    \mathbf{J} = \nabla \psi \times \nabla k,
\end{equation}
where $k$ is the current field line label, which is expanded in powers of $\epsilon\rho$ similar to $\psi$ and $\alpha$. To obtain the lowest order coefficients of $\alpha$ and $k$, we start by noting that, to lowest order in $\epsilon$, the magnetic field and the current density vectors are tangent to the magnetic axis, yielding

\begin{equation}\label{on-axis B}
    \mathbf{B} = B_0(s)\mathbf{e}_s + O(\epsilon),
\end{equation}
and
\begin{equation}\label{on-axis J}
    \mathbf{J} = J_0(s)\mathbf{e}_s + O(\epsilon),
\end{equation}
where $J_0$ is the on-axis current density.

We now set the lowest order components of $\mathbf{B}$ and $\mathbf{J}$ equal to their equivalent Clebsch representations, yielding
\begin{equation}\label{alpha 0 omega}
    \dot{\alpha_0} = \frac{B_0}{2\psi_2},
\end{equation}
and
\begin{equation}\label{k 0 omega}
    \dot{k_0} = \frac{J_0}{2\psi_2},
\end{equation}
where dots represent derivatives with respect to $\omega$. Similarly, setting the lowest order component of $\mathbf{J} = \nabla \times \mathbf{B}$ equal to Eq. \eqref{on-axis J}, we find that
\begin{equation}\label{alpha 0 s}
   \alpha_0' = -\frac{B_0' \dot{\psi_2}+2 J_0 \psi_2- B_0 \dot{\psi_2}'}{2
   \ddot{\psi_2} \psi_2+8 \psi_2^2},
\end{equation}
where primes denote derivatives with respect to $s$.

Next, we write the fields $\mathbf{B}$ and $\mathbf{J}$ in the ideal MHD equation Eq. \eqref{ideal MHD} using the Clebsch representation together with Eqs. \eqref{alpha 0 omega} and \eqref{k 0 omega}. This allows us to find a set of equations for $k_0'$ and $\alpha_1$ needed to find the on-axis magnetic well. Additionally, we express the pressure gradient as $\nabla p = p'(\psi) \nabla \psi$, where $p'(\psi)$ is expanded as $p'(\psi) \simeq p_2$ and it is assumed that the higher order terms are small. This yields
\begin{equation}\label{k 0 s}
     k_0' = \frac{-p_2 + J_0\alpha_0'}{B_0}.
\end{equation}
The same procedure using the expression $\mathbf{J} = \nabla \times \mathbf{B}$ instead of the Clebsch representation for $\mathbf{J}$ yields
\begin{equation}
    2\frac{\partial}{\partial \omega}\left(\frac{\alpha_1}{\sqrt{\psi_2}}\right)\psi_2^{3/2} = \kappa\cos\theta B_0 - \frac{3}{2}\frac{B_0\psi_3}{\psi_2}.
\end{equation}
We note that since the on-axis magnetic field and current density are parallel to the axis they are related by $J_0 = \lambda B_0$ for some $\lambda = \lambda(s)$. Furthermore, since $\nabla \cdot \mathbf{J} = 0$, this implies that $\partial \lambda/\partial s = 0$, which shows that $\lambda$ is a constant. Consequently, from the Eq. \eqref{k 0 s}, we find that $k_0 = \lambda \alpha_0 - p_2\int ds/B_0$.

\section{Magnetic Well}\label{sec:magnetic well}
The magnetic well, written as $V''(\psi)$, is a commonly employed metric for stability against interchange modes in magnetic confinement fusion devices. A more detailed analysis of the role of $V''(\psi)$ in Mercier's criterion is discussed in Appendix \ref{appB}. To derive an expression for the magnetic well in Mercier's coordiantes, we start by expressing the volume $V$ enclosed by a toroidal surface in $(\rho,\theta,s)$ coordinate system as
\begin{equation}\label{pws_vol}
    V = \int_0^\rho\int_0^{2\pi}\int_0^L \frac{d\rho d\theta ds}{\nabla\rho\times\nabla\theta\cdot\nabla s},
\end{equation}%
where $\nabla\rho\times\nabla\theta\cdot\nabla s = 1/[\rho(1-\kappa\rho\cos\theta)]$ is the inverse of the Jacobian. Assuming the existence of nested flux surfaces, we can express $V$ in terms of $\psi$ by moving from Mercier's coordinates to $(\psi,\theta,s)$ coordinates, yielding
\begin{equation}\label{magnetic well integral}
    V = \int_0^\psi\int_0^{2\pi}\int_0^L d\psi d\theta ds \frac{\rho(1-\kappa\rho\cos \theta)}{\partial\psi/\partial\rho}.
\end{equation}%
The magnetic well can then be written as
\begin{equation}\label{pre-substitution V''}
    V''(\psi) = \int_0^{2\pi}\int_0^L d\theta ds \frac{\partial}{\partial\psi}\left[ \frac{\rho(1-\kappa\rho\cos \theta)}{\partial\psi/\partial\rho}\right].
\end{equation}%

We now compute the lowest order component of the integrand in Eq. \eqref{magnetic well integral}. Starting with the power series of $\psi$ in terms of $\rho$
\begin{equation}
    \psi = \psi_2\rho^2 + \psi_3\rho^3 + \psi_4\rho^4 + O(\rho^5),
\end{equation}
We can then find the inverse series for $\rho$, yielding
\begin{equation}\label{fourth_order_expansion}
    \rho = \frac{\psi^{\frac{1}{2}}}{(\psi_2)^{\frac{1}{2}}} - \frac{\psi_3\psi}{2(\psi_2)^{2}} + \frac{[5(\psi_3)^2-4\psi_2\psi_4]\psi^{3/2}}{8(\psi_2)^{7/2}} + O(\psi^2).
\end{equation}
We then substitute Eq. \eqref{fourth_order_expansion} into Eq. \eqref{pre-substitution V''}. Taylor expanding the denominator of Eq. \eqref{pre-substitution V''}, simplifying, and keeping only the terms linear with $\psi$ yield the lowest order magnetic well
\begin{equation}\label{magnetic well}
    V''(0) = \int_0^L \int_0^{2\pi} d\theta ds \left[\frac{3(\psi_3)^2}{2(\psi_2)^4} + \frac{\kappa\psi_3\cos \theta}{(\psi_2)^3} - \frac{\psi_4}{(\psi_2)^3}\right].
\end{equation}
This expression is equivalent to Eq. (56) in Chapter 3 of \citet{Mercier1974LecturesConfigurations}. However, we note that \citet{Mercier1974LecturesConfigurations} denotes the poloidal flux as $\psi$, which yields extra conversion factors between the toroidal and poloidal fluxes described in the same chapter.

In order to simplify Eq. \eqref{magnetic well} and rewrite it in terms of $\psi_2$ and $\psi_3$ only, we subtract the $\rho$ component of the ideal MHD equation in the form $\mathbf{J} \times \mathbf{B} - \nabla p = 0$ from Eq. \eqref{magnetic well} together with the total derivatives $\partial/\partial\omega(\alpha_2/B_0\psi_2)$, $-\partial/\partial s(\alpha_0'\dot{\psi_2}/4B_0\psi_2^2)$, $-\partial/\partial\omega (\alpha_1\psi_3/2B_0\psi_2^2)$, and $\partial/\partial s(\psi_2'/\psi_2^3)$. Since these terms are derivatives of functions that are either periodic over $\theta$ or $s$, their integrals over $\theta$ and $s$ in Eq. \eqref{magnetic well} vanish. Using the integral formulas derived in Appendix \ref{appA} to perform the integration over $\theta$, our expression for the magnetic well reads
\begin{align}\label{final well}
    V''(0) &= \int_0^L \frac{ds}{4\pi^2B_0^4\left(1-\mu^2\right)^{5/2}}\left[3\pi\left(-1+\mu^2\right)^2B_0'^2 \right.\nonumber\\
    &\left.+\pi B_0^2\left[-\left(-1+\mu^2\right)^2\left(-\lambda^2 + 2\kappa^2 \left(1-\mu\cos 2\delta\right) + \mu^2\left[\lambda^2 - 4\left(\tau^2 - \delta'^2\right)\right] \right) \right.\right.\nonumber\\
    &\left.\left.+ \mu'^2\right] + 4B_0\left(-1 + \mu^2\right)\left(-\sqrt{-1-\mu^2}\left(\pi^2p_2 + 2\kappa\psi_{3\delta}  \right) \right.\right.\nonumber\\
    &\left.\left.+ \mu^2\psi_{3\kappa}\sqrt{1-\mu^2} + \mu\left[3\kappa \sqrt{1-\mu^2}\left(\psi_c\cos\delta + \psi_s\sin\delta \right) + \pi B_0' \mu'\right]  \right)\right],
\end{align}
where $\psi_c = \psi^c_{31} + \psi^c_{33}$, $\psi_s = -\psi^s_{31} + \psi^s_{33}$, $\psi_{3\delta} = \psi_{31}^c \cos \delta+\psi_{31}^s \sin\delta$, and $\psi_{3\kappa} = -\kappa\left[ (\psi_{31}^c+3 \psi_{33}^c)\cos \delta- (\psi_{31}^s-3 \psi_{33}^s)\sin \delta\right] +\pi ^2 p_2$.

We note that $\psi_3$ and $\psi_4$, which are are higher order terms in the near-axis expansion, both appear in Eq. \eqref{magnetic well}. However, only $\psi_3$ is needed in the final expression for the magnetic well as $\psi_4$ can be expressed in terms of $\psi_2$ and $\psi_3$ using the ideal MHD equation Eq. \eqref{ideal MHD}. As shown in \citet{Jorge2020Near-axisAxis}, the differential equation for $\psi_3$ has source terms proportional to the lower order quantities $\mu$, $\delta$, and $\psi_{20}$. Therefore, we conclude that although lower order $\psi_2$ components are sufficient to describe the surfaces near the axis, we need higher order $\psi_3$ terms in order to accurately compute the magnetic well.

Finally, we mention here a separate way of deriving the magnetic well for the vacuum case, i.e., when the magnetic field is obtained using Laplace's equation, \cref{laplace}, and the surface shapes obtained using $\nabla \phi \cdot \nabla \psi = 0$.
While the on-axis magnetic well is still given by \cref{magnetic well} for both vacuum and non-vacuum cases, the term proportional to $\psi_4$ is cast in terms of $\psi_2$ and $\psi_3$ differently, namely, using the definition for the toroidal flux in \cref{eq:torflux}.
Indeed, replacing the magnetic field with $\mathbf B = \nabla \phi$ with \citep{Jorge2020ConstructionApproach,Jorge2020Near-axisAxis}
\begin{align}
    \phi = \int B_0 ds + \rho^2\frac{B_0}{2}\left[(\ln B_0^{-1/2})'+\mu u' \sin 2u-\frac{\eta'}{2}\cos 2 u\right]+O(\epsilon^3),
\end{align}
and $\mu=\tanh \eta$, the third order component of \cref{magnetic well} yields
\begin{align}
    \frac{3(\psi_3)^2}{2(\psi_2)^4} + \frac{\kappa\psi_3\cos \theta}{(\psi_2)^3} - \frac{\psi_4}{(\psi_2)^3}&=\frac{1}{8 B_0 (\psi_2)^2}\left[B_0'(2 \mu \sin 2 u (\delta''-\tau)-\eta' \cos 2u)+4 \cos \theta^2 \kappa^2\right.\nonumber\\
    &\left.-B_0''+B_0(2 \sin 2u[(2-\mu^2)\eta'(\delta'-\tau)+\mu(\delta''-\tau')]\right.\nonumber\\
    &\left.+\cos 2u [4\mu (\tau - \delta')^2-\eta''] )\right]+2\frac{\kappa\psi_3\cos \theta}{(\psi_2)^3}.
\label{eq:vacwell1}
\end{align}
Using the integration rules derived in \cref{appA}, we can integrate \cref{eq:vacwell1} over $s$ and $\theta$, retrieving the vacuum ($\lambda=p_2=0$) form of the magnetic well expression in \cref{final well}.

We note that, with respect to \citet{Solovev1970PlasmaSystems}, we have generalized the formula for the magnetic well to the realistic case of finite current density. Additionally, we have derived the formula for the magnetic well in terms of surface quantities $\psi_2$ and $\psi_3$, while the final criterion derived in \citet{Mercier1974LecturesConfigurations} does not isolate the magnetic well and is not explicitly written in terms of surface quantities.

\section{Mercier's Criterion}\label{sec:mercier's criterion}
Mercier's criterion states that the plasma will be approximately stable against interchange perturbations around rational surfaces if the following criterion is satisfied
\begin{equation}\label{Full criterion}
    \left[\frac{1}{2}\frac{d\iota}{d\psi} + \int \frac{\mathbf{B}\cdot\mathbf{\Xi} dS}{|\nabla \psi|^3}\right]^2 + \int \frac{B^2 dS}{|\nabla \psi|^3}\left[\frac{dp}{d\psi}\frac{d^2V}{d\psi^2}-\int\frac{|\mathbf{\Xi}|^2 dS}{|\nabla \psi|^3}\right] \geq 0,
\end{equation}
where $\iota$ is the rotational transform, $I = \left(1/L\right)\left[\int (\mathbf{J} \cdot \nabla s) dV\right]$ is the total toroidal current, and
\begin{equation}
    \mathbf{\Xi} = \mathbf{J} - \frac{dI}{d\psi}\mathbf{B},
\end{equation}
with $I'(\psi) = \lambda + O(\epsilon)$. As discussed in Appendix \ref{appB}, to lowest order, the Mercier criterion reduces to the sum of the magnetic well term and the surface integral of $\Xi^2$
\begin{equation}
    p_2 V'' - \int \frac{|\mathbf \Xi|^2 dS}{|\nabla \psi|^3}\ge 0.
\end{equation}

We first note that the zeroth order $\mathbf \Xi$ vector vanishes as $\mathbf{\Xi}_0 = (J_0 - \lambda B_0)\mathbf{e}_s = 0$.
In order to simplify its first order component and write it using Mercier's coordinates, since $\nabla \cdot \mathbf{\Xi} = 0$, we rewrite $\mathbf{\Xi}$ as 
\begin{equation}
    \mathbf{\Xi} = \nabla \psi \times \nabla G,
\end{equation}
where $G$ is defined as
\begin{equation}
    G = k - I'(\psi)\alpha.
\end{equation}
We additionally expand $G$ as a power series in $\epsilon\rho$ similar to Eq. \eqref{B field expansion}. To lowest order, we find that $G_0 = -p_2\int ds/B_0$. To first order, $G_1$ is given by
\begin{equation}
    G_1 = k_1 - \lambda\alpha_1.
\end{equation}
We use as an ansantz for $G_1$ the following expression \citep{Mercier1974LecturesConfigurations}
\begin{equation}\label{ansatz}
    G_1= \sqrt{B_0}\left[G_{1c}(s)e^{\eta/2}\cos u+ G_{1s}(s)e^{-\eta/2}\sin u\right],
\end{equation}
where $\mu = \tanh \eta$.
We note that the parameters $G_{1c}$ and $G_{1s}$ are related to the first order parallel plasma current $J_{1s}$ via
\begin{equation}
    J_{1s}=-\kappa \lambda \cos \theta+\frac{2\pi B_0^{3/2}}{\sqrt{1-\mu^2}}\left[(1-\mu)e^{\eta/2} G_{1c} \sin u - (1+\mu)e^{-\eta/2} G_{1s} \cos u\right].
\end{equation}

Substituting Eq. \eqref{ansatz} into the $\omega$ component of the ideal MHD equation and equating the resulting coefficients of the $\cos u$ and $\sin u$ term, we obtain the first order ordinary differential equations for $G'_{1c}(s)$ and $G'_{1s}(s)$
\begin{equation}\label{G1c}
    G'_{1c} = \frac{2 p_2 \kappa e^{-\frac{\eta}{2}} \cos \delta }{B_0^{3/2}}-\frac{1}{2} G_{1s} \sech \eta
   \left(\lambda +2 \delta'-2 \tau\right),
\end{equation}
and
\begin{equation}\label{G1s}
    G'_{1s} = \frac{2 p_2 \kappa e^{\frac{\eta}{2}} \sin \delta}{B_0^{3/2}}+\frac{1}{2} G_{1c} \sech\eta
   \left(\lambda +2 \delta'-2 \tau\right).
\end{equation}
We find that the same differential equations as Eqs. \eqref{G1c} and \eqref{G1s} are obtained by equating the $\cos 3u$ and $\sin 3u$. This suggests that the third harmonics do not yield any additional information, and our ansantz is indeed a solution of $\mathbf{J} \times \mathbf{B} = \nabla p$.
The first order solution of $\mathbf{\Xi}$ can then be written as
\begin{equation}\label{Xi rho}
    \Xi_\rho = 4\pi p_2 \cos u \sin u \sinh \eta,
\end{equation}
\begin{equation}\label{Xi omega}
    \Xi_\omega = 2\pi p_2 (\cosh \eta + \cos 2u \sinh \eta),
\end{equation}
and
\begin{equation}\label{Xi s}
    \Xi_s = 2 \pi  B_0^{3/2} e^{-\frac{\eta}{2}} G_{1cs},
\end{equation}
where $G_{1cs} = e^{\eta} G_{1s} \cos u-G_{1c} \sin u$. The resulting $\Xi^2$ quantity entering Mercier's criterion is given by
\begin{equation}
    \Xi^2 = 4 \pi ^2 e^{-2 \eta} \left[B_0^3 e^{\eta} G_{1cs}^2+p_2^2 \left(e^{4 \eta} \cos ^2 u\sin ^2 u\right)\right].
\end{equation}
Noting that $dS = |\nabla \psi|\sqrt{g} d\theta ds$, the surface integral of $\Xi^2$ in Eq. \eqref{Full criterion} can be written as
\begin{equation}\label{xi squared integral}
   \int_s \frac{|\mathbf{\Xi}|^2 dS}{|\nabla \psi|^3} = \int_0^L\int_0^{2\pi} d\theta ds \frac{\Xi^2 \rho(1-\kappa\rho\cos\theta)}{\partial\psi/\partial\rho |\nabla \psi|^2} = \int_0^L ds \left(\frac{p_2^2}{B_0^3} + \frac{e^{\eta}G_{1c}^2 + G_{1s}^2}{1 + e^{\eta}}\right) + O(\epsilon).
\end{equation}
The quantities in Eq. \eqref{xi squared integral} can be evaluated for a particular toroidal current $\lambda$, pressure gradient $p_2$, and elliptical surface $\psi_2$ by solving the system of equations in Eqs. \eqref{G1c} and \eqref{G1s}. We note that when we set $p_2 = 0$, solving Eqs. \eqref{G1c} and \eqref{G1s} with periodic boundary conditions yields $G_{1c} = G_{1s} = 0$. Therefore, in the limit of vanishing pressure gradient, Eq. \eqref{xi squared integral} evaluates to $0$, and the magnetic well becomes the dominant term in the criterion.

Finally, following similar reasoning, but using the on-axis magnetic field, it can be shown that, to lowest order,
\begin{equation}\label{Bsquared integral}
    \int \frac{B^2 dS}{|\nabla \psi|^3} = \frac{L}{4\pi}.
\end{equation}
\section{Tokamak with a Circular Cross Section}\label{sec:circular cross section}
We now apply Eq. \eqref{final well} for the case of a tokamak with a circular cross section. The resulting expression is compared with its counterpart in the inverse coordinate approach as derived in \citet{Landreman2020MagneticAxis}.

\subsection{Direct Approach}
For an axisymmetric design with a circular cross section, the quantities $\mu$, $\delta$, and $\tau$, together with the derivatives with respect to s (primed quantities) in Eq. \eqref{final well} are 0. However, $\psi_3$ may still be finite due to the presence of a Shafranov shift. An equation for $\psi_3$ is obtained by first noting that, in this case, Eqs. \eqref{G1c} and \eqref{G1s} simplify to 
\begin{equation}\label{G1sAxi}
    G_{1s} = \frac{4p_2\kappa}{\lambda B_0^{3/2}},
\end{equation}
and
\begin{equation}\label{G1cAxi}
    G_{1c} = 0,
\end{equation}
respectively. Plugging  Eqs. \eqref{G1sAxi} and \eqref{G1cAxi} into the the Clebsch representation of $\mathbf{J}$, and equating its $\mathbf{e}_s$ component to $\mathbf{J} = \nabla \times \mathbf{B}$, we find that
\begin{equation}\label{psi31cAxi}
    \psi^c_{31} = \frac{\kappa}{4}\left(\frac{16\pi p_2}{\lambda^2} - B_0\right),
\end{equation}
together with $\psi^s_{31} = \psi^c_{33} = \psi^s_{33} = 0$. Using Eqs. \eqref{alpha 0 omega} and \eqref{alpha 0 s}, in addition to Eq. (5.6) of \citet{Jorge2020Near-axisAxis}, it can also be shown that in axisymmetry with circular cross sections, $\lambda = 2\iota_0\kappa$ with $\iota_0$ the rotational transform on-axis. Substituting this expression and Eq. \eqref{psi31cAxi} into Eq. \eqref{final well}, integrating over a circular axis with $\kappa = 1/R$, where R is the major radius, and taking a low $\beta$ limit so that the remaining terms proportional to $p_2$ can be neglected, yields
\begin{equation}\label{wellAxi}
    V'' = \frac{2(-1+\iota^2)}{B_0^2R}.
\end{equation}
Therefore, this configuration is stable if $\iota < 1$. For tokamaks, the safety factor $q = 1/\iota$ is more commonly used, recovering the Kruskal-Shafranov limit $q > 1$ \citep{Freidberg2014IdealMHD}.

\subsection{Inverse Approach}
The magnetic well in the inverse approach for quasisymmetry reads \citep{Landreman2020MagneticAxis}
\begin{align}
\label{eq:wellQS}
V''= \frac{|G_0|}{B_0^3}
\left(3 \overline{\eta}^2 - \frac{4 B_{20}}{B_0} - \frac{2 p_{2I}}{B_0^2}\right),
\end{align}
where, in axisymmetry (which is a special case of quasisymmetry), $G_0=B_0R$.
In axisymmetry, $B_{20}$ is found from Eqs. (A27) and (A28) of \citet{Garren1991ExistenceStellarators}, and can be written as
\begin{align}
    \frac{B_{20}}{B_0}&=-\frac{p_{2I}}{B_0^2}+\frac{1}{3-\overline \eta^4 R_{00}^4+3\sigma^2}\left[-\frac{p_{2I}F^2}{2 I_2^2 R_{00}^2}+3(\overline \eta^4 R_{00}^4-1-3\sigma^2)\frac{B_{2c}}{B_0}\right.\nonumber\\
    &\left.+6\sigma(\overline \eta ^4 R_{00}^4+\sigma^2)\frac{B_{2s}}{B_0}+\frac{\overline \eta^2}{2}(7-2\overline \eta^4 R_{00}^4+4 \sigma^2)+\frac{4 I_2^2 \overline \eta^6 R_{00}^6(F-3)}{B_0^2 F^2}\right],
\end{align}
where $F=\overline \eta^4 R_{00}^4 + \sigma^2+1$, $I_2 = \iota_0B_0\kappa$, and $p_{2I} = B_0p_2/2$. Furthermore, in up-down symmetric geometries, $\sigma = 0$, and for circular cross sections, $\overline \eta=1/R$. Finally, in the low $\beta$ limit, $B_{20}/B_0$ reduces to
\begin{equation}
    \frac{B_{20}}{B_0} = \frac{5\overline{\eta}^2}{4} - \frac{I_2^2}{2B_0^2}.
\end{equation}
The magnetic well is then given by
\begin{equation}
    V'' = \frac{2(-1+\iota^2)}{B_0^2R},
\end{equation}
which is equivalent to Eq. \eqref{wellAxi}.
\section{Numerical Results}\label{sec:numerical verification}
Numerical verification for a near-axis analytical expression for Mercier's criterion and the magnetic well has only been done very recently using the inverse coordinate approach \citep{Landreman2020MagneticAxis}, albeit for only quasisymmetric devices. A numerical study is a crucial step needed to verify the formula for Mercier's criterion on-axis. In this work, we perform a preliminary analysis for the magnetic well and the lowest order criterion. We note that, near the axis of low $\beta$ systems, the magnetic well becomes the dominant term. Indeed, previous stellarator optimization studies have mainly focused on the magnetic well as a proxy for MHD stability \citep{Beidler1990, Anderson1995TheStatus, Drevlak2019OptimisationROSE}. However, we numerically verify the lowest order Mercier's criterion for a case where the $\Xi^2$ term is also significant.

To verify the expression for the magnetic well, Eq. \eqref{final well}, we apply it to the optimized stellarator design of Wendelstein 7-X, as well as two quasisymmetric stellarators from \citet{Landreman2019ConstructingOrder}. In particular, for W7-X we choose the A configuration from \citet{Geiger2015PhysicsW7-X} in vacuum. For the quasisymmetric stellarators, we choose the quasi-axisymmetric hybrid tokamak-stellarator of Section 5.3 of \citet{Landreman2020MagneticAxis} in finite $\beta$ and the quasi-helical symmetric stellarator of Section 5.4 in vacuum. For both quasisymmetric configurations, we specifically choose to focus on a surface with $\psi \sim 0.002$ T m\textsuperscript{2}, $\Bar{B} = 1$ T, and $R = 1$ m, rendering the expansion parameter $\epsilon\rho \sim \sqrt{\psi/\Bar{B}R^2} \sim 0.04$. For W7-X, we focus on a surface with $\psi \sim 0.001$ T m\textsuperscript{2}, $\Bar{B} \sim 3 \ T$, and $R \sim 5.5 \ m$, rendering $\epsilon\rho \sim 0.01$.

To generate the equilibria, we use the VMEC code \citep{Hirshman1983Steepest-descentEquilibria}. VMEC uses a cylindrical coordinate system, with the position vector given by
\begin{equation}
    \mathbf{r} = R\mathbf{e}_R(\Phi) + Z\mathbf{e}_Z,
\end{equation}%
where $(R,\Phi,Z)$ are the standard cylindrical coordinates and $(\mathbf{e}_R,\mathbf{e}_\Phi,\mathbf{e}_Z)$ the corresponding unit basis vectors. VMEC surfaces are parameterized by a poloidal angle $\theta_v$ and the standard toroidal angle $\Phi$. In a stellarator-symmetric configuration such as the ones used here, $R$ and $Z$ can be written as
\begin{equation}
    R = \sum_{m,n} R_{mn}\cos(m\theta_v - n\Phi),
\end{equation}%
and
\begin{equation}
    Z = \sum_{m,n} Z_{mn}\sin(m\theta_v - n\Phi).
\end{equation}%
The magnetic axis is parameterized via $\Phi_a$, with $0 \leq \Phi_a < 2\pi$. For W7-X, we write its magnetic axis as %
\begin{equation}
    \mathbf{r}_0(\Phi_a) = [5.56 + 0.37\cos(5\Phi_a) + 0.02\cos(10\Phi_a)]\mathbf{e}_R - [0.31\sin(5\Phi_a) + 0.02\sin(10\Phi_a)] \mathbf{e}_Z,
\end{equation}%
which is the same axis as in Eq. 7.4 in \citet{Jorge2020ConstructionApproach}.
The magnetic axis of the quasi-axisymmetric and quasi-helical symmetric stellarators are
\begin{equation}
    \mathbf{r}_0(\Phi_a) = [1 + 0.09\cos(2\Phi_a)]\mathbf{e}_R - [0.09\sin(2\Phi_a)]\mathbf{e}_Z,
\end{equation}
and
\begin{align}\label{qh}
    \mathbf{r}_0(\Phi_a) &= [1 + 0.17\cos(4\Phi_a) + 0.01804\cos(8\Phi_a) + 0.001409\cos(12\Phi_a)]\mathbf{e}_R \nonumber \\&- [0.1583\cos(4\Phi_a) + 0.01820\cos(8\Phi_a) + 0.001548\cos(12\Phi_a)]\mathbf{e}_Z,
\end{align}
respectively. We note that terms in Eq. \eqref{qh} with coefficients less than $10^{-3}$ are neglected in this text, but are included in the numerical calculations. Finally, we perform a change of variables from $s$ in Mercier's coordinates to $\Phi_a$ in VMEC coordinates using the fact that $ds/d\Phi_a = \lvert d\mathbf{r}_0/d\Phi_a\rvert$.

Using the SENAC code \citep{SENAC}, we perform a nonlinear least-squares fit to derive the Fourier coefficients for $\mu$, $\delta$, $B_0$, and $\psi_3$. Using these coefficients, we write the magnetic field on-axis as $B_0 = \sum_{n} B_{0n}\cos(nN_{fp}\Phi_a)$, $\mu = \sum_{n} \mu_n\cos(nN_{fp}\Phi_a)$, and $\delta = -N_{fp}\Phi_a/2 + \sum_{n} \delta_n\sin(nN_{fp}\Phi_a)$, with $N_{fp}$ the number of field periods of the device. For W7-X, $N_{fp} = 5$. For the quasi-axisymmetric and quasi-helical symmetric stellarators, $N_{fp} = $ 2 and 4, respectively. To satisfy stellarator symmetry, $\psi_3$ is written as
\begin{equation}
    \psi_3 = \sum_{m,n} \psi_3^{mn}\cos(m\theta - nN_{fp}\Phi_a),
\end{equation}
which can be expanded to calculate $\psi_{31}^c$, $\psi_{31}^s$, $\psi_{33}^c$, and $\psi_{33}^s$. The Fourier coefficients for these functions obtained from the W7-X fit are shown in Table. \ref{table:fourier coeff}. We note that a fitting algorithm is used here because we are using equilibria generated by VMEC. Since VMEC uses a cylindrical coordinate system to represent the boundary surface in a discrete set of points, there is no exact elliptical or triangular representation of the surface. Therefore, we need a fitting algorithm to an elliptical or triangular function to calculate values of $\mu$ or $\psi_3$.

\begin{table}
\centering
\begin{tabular}{l|lllll}
n  & 0      & 1      & 2     & 3     & 4    \\ \hline
$B_0$ & 2.75 & 0.07 & -0.04 & -0.05 & -0.05 \\
$\delta$ & - & 0.56  & -0.12 & 0.03 & - \\
$\mu$ & 0.68 & 0.19 & -0.04 & - & - \\
$\psi_3^{1n}$ & -0.42 & -0.51 & 0.60 & 0.31 & -0.18  \\
$\psi_3^{3n}$ & 0.83  & 0.35  & 0.10 & 0.69 & -0.24 
\end{tabular}
\caption{\label{tab:w7xfit} Fitting results of the W7-X surface $\psi \sim 0.01$ T m$^{-2}$ to the values in Eq. \eqref{final well}. Only the parameters with absolute value greater than $0.01$ are shown.}
\label{table:fourier coeff}
\end{table}

\begin{figure}
    \centering
    \includegraphics[scale=0.75]{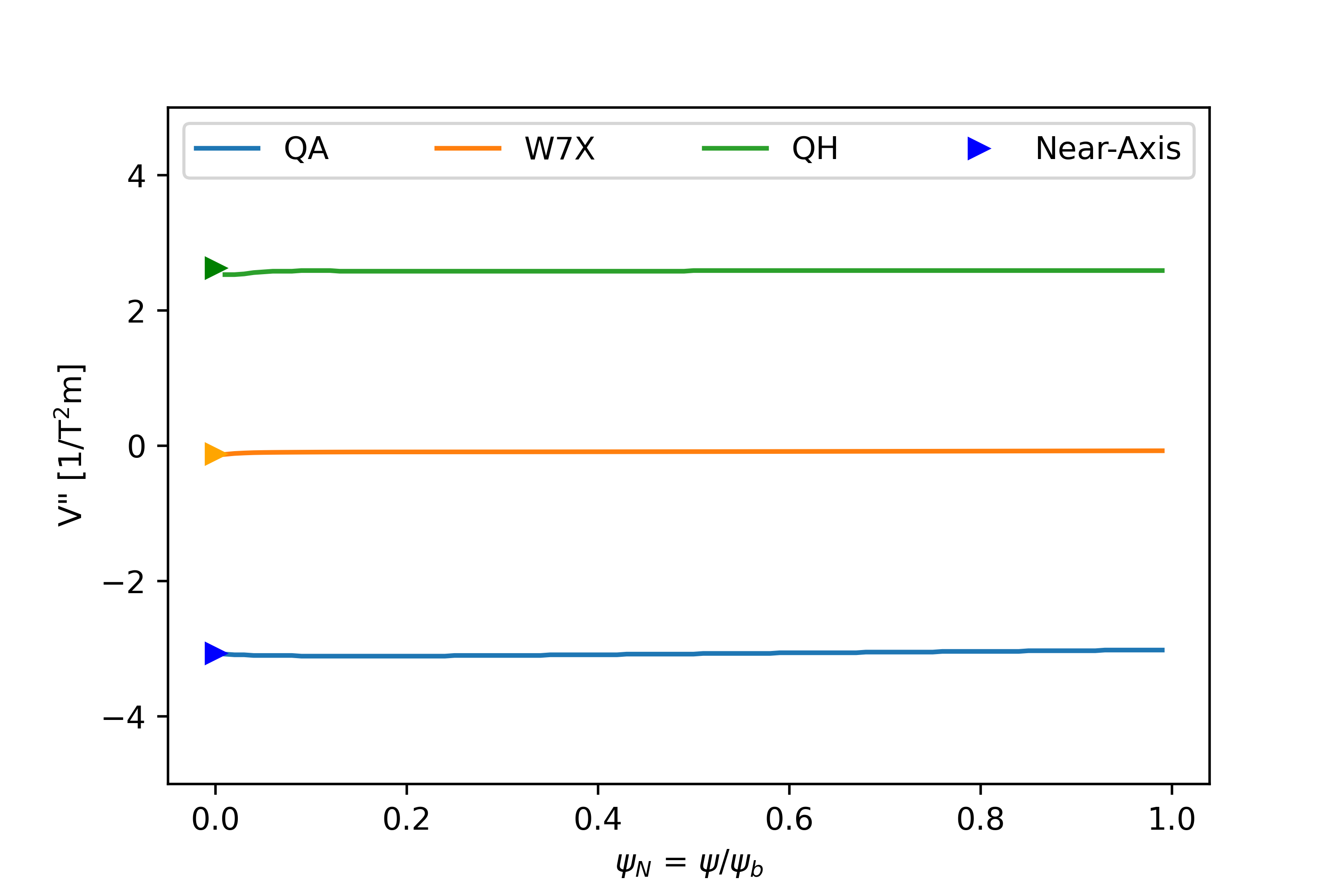}
    \caption{The magnetic wells calculated from the Eq. \eqref{final well} and VMEC for a quasi-axisymmetric, quasi-helical symmetric and W7-X stellarators with $\psi_b$ the value of $\psi$ at the plasma boundary. The quasisymmetric stellarators are fitted at the boundary, while W7-X is fitted at its innermost surface.}
    \label{fig:magnetic_well}
\end{figure}

The resulting magnetic wells as a function of the normalized toroidal flux $\psi_N$ calculated for the quasisymmetric stellarators and W7-X are shown in Fig. \ref{fig:magnetic_well}. In particular, the lines shown are the magnetic wells calculated using VMEC. Since VMEC uses a uniform radial grid in terms of $\psi$, it is typically most accurate at the boundary surface. Due to the fact that the quasisymmetric stellarators have small expansion parameters at the boundary, we choose to make fits at the boundary to calculate the magnetic well. Since the boundary surface is near the axis, there is little variation of the magnetic well. However, since the W7-X VMEC file has a boundary surface far from the axis, we choose the innermost surface. In all cases, we calculate a near-axis magnetic well within a few hundredths in magnitude of the VMEC magnetic well, thereby verifying Eq. \eqref{final well} for both quasisymmetric and arbitrary stellarator designs.

\begin{figure}
    \centering
    \includegraphics[scale=0.075]{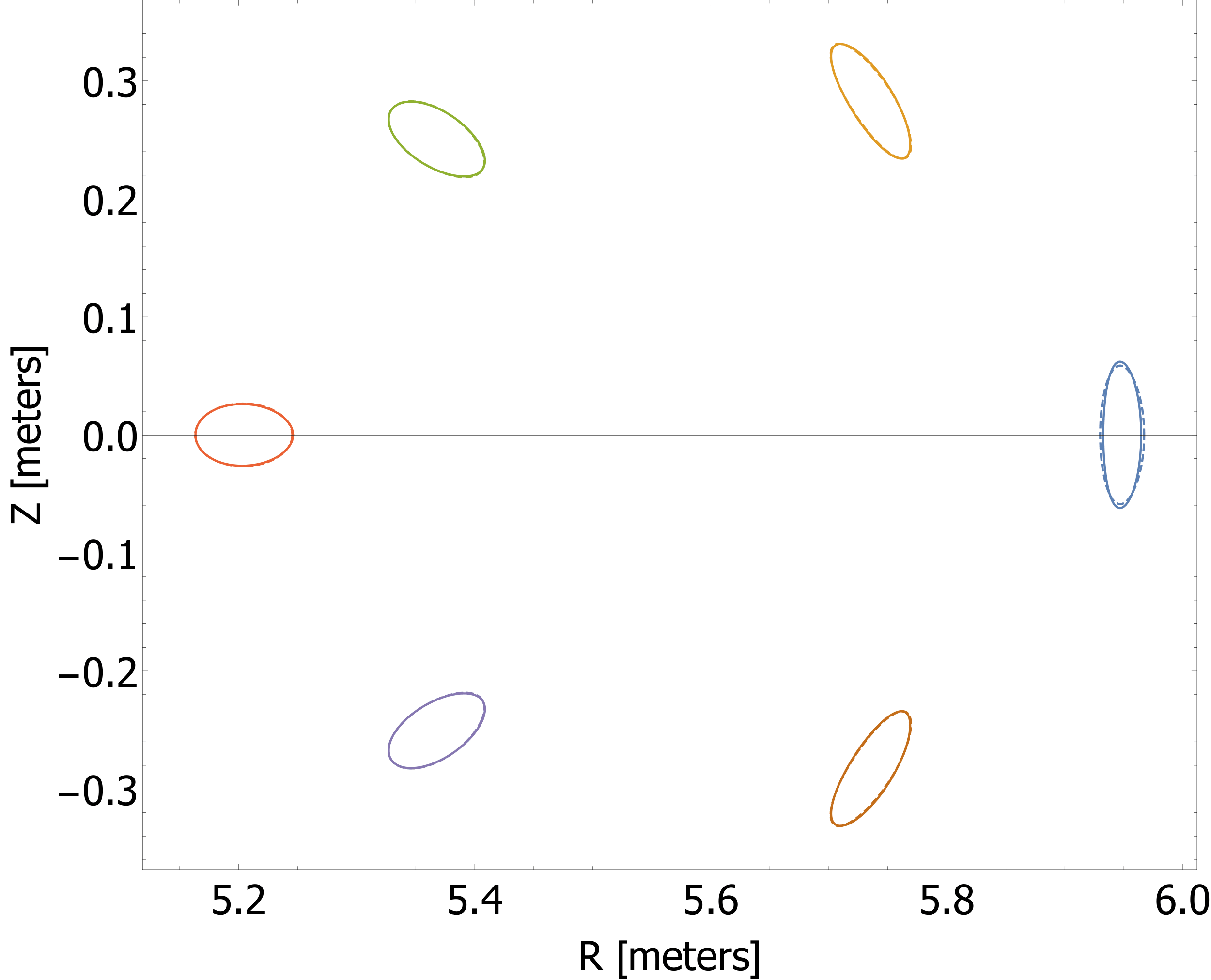}
    \caption{The poloidal cross sections of W7-X. There are dashed and solid lines for the near-axis and VMEC cross sections respectively, but the differences are nearly indistinguishable.}
    \label{fig:w7x poloidal}
\end{figure}

\begin{figure}
\begin{center}
    \begin{subfigure}{0.45\textwidth}
\includegraphics[width=0.9\linewidth, height=5cm]{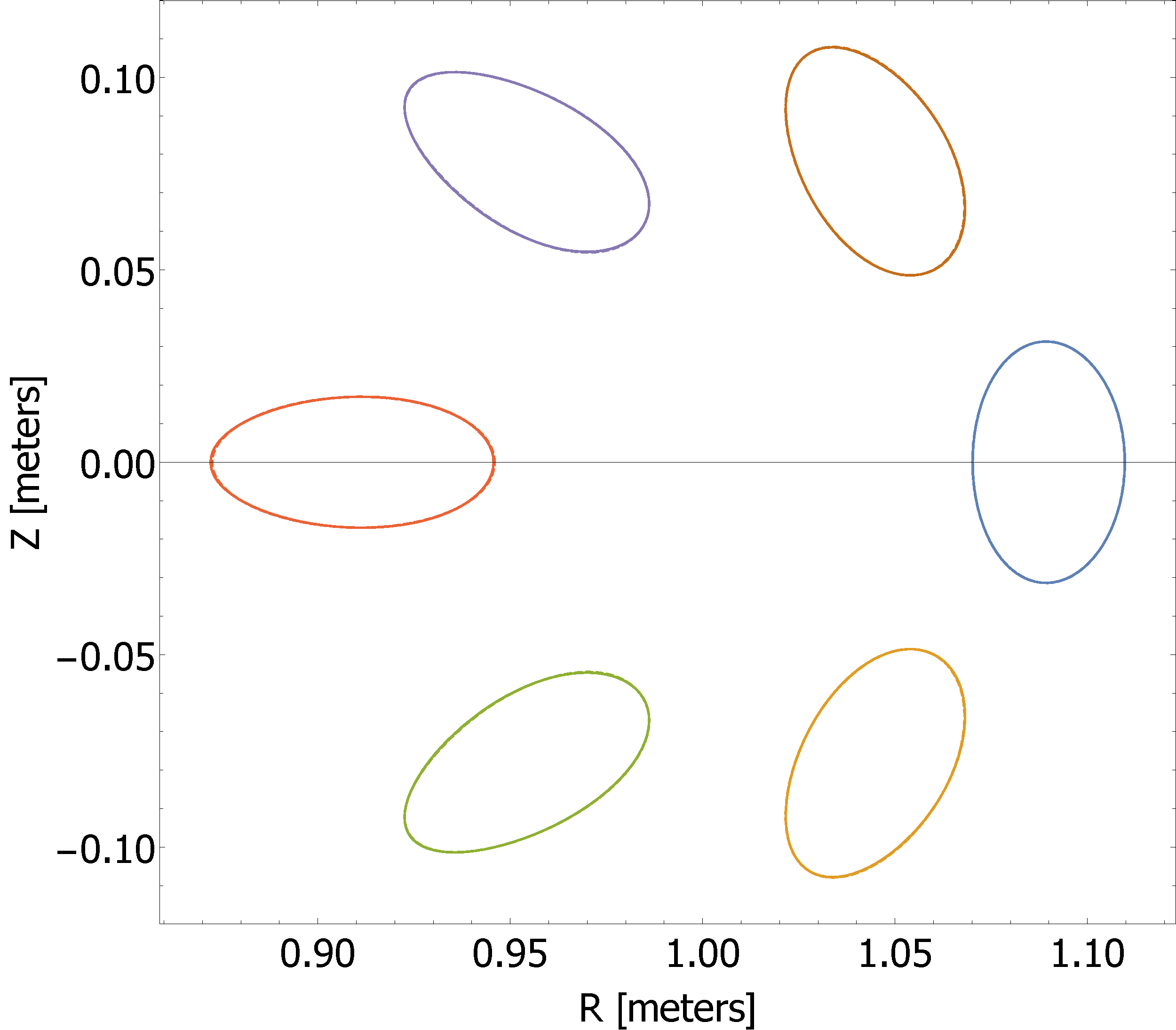} 
\end{subfigure}
\begin{subfigure}{0.45\textwidth}
\includegraphics[width=0.9\linewidth, height=5cm]{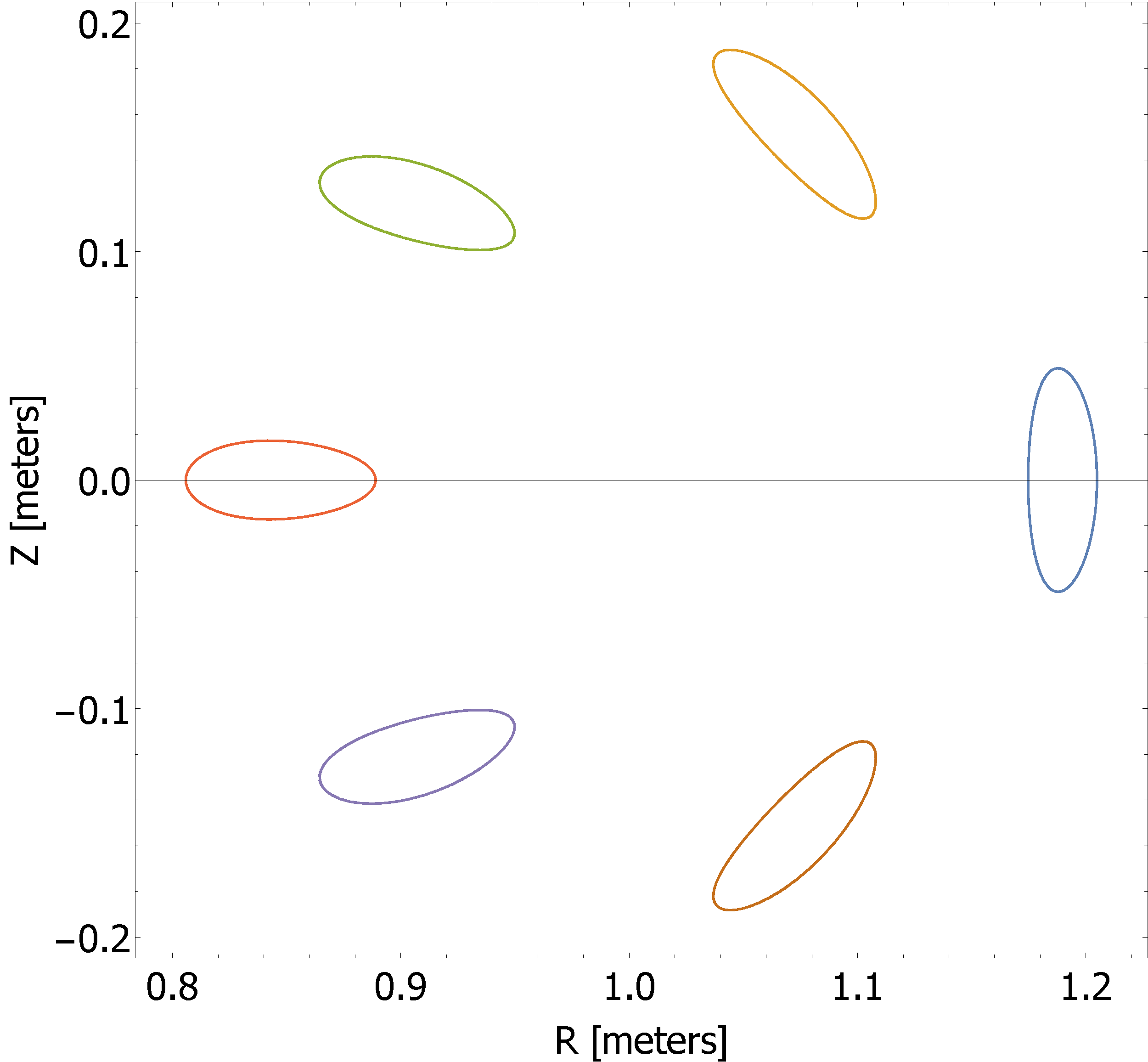}
\end{subfigure}
\end{center}

\caption{Left: The poloidal cross sections of the fitted quasi-axisymmetric stellarator. Right: The poloidal cross sections of the fitted quasi-helically symmetric stellarator. There are dashed and solid lines for the near-axis and VMEC cross sections respectively, but the differences are nearly indistinguishable.}
\label{fig:poloidal plots}
\end{figure}

The fit's accuracy is evident from the plot of the poloidal cross sections. As seen in Fig. \ref{fig:w7x poloidal}, there are only very slight differences between the VMEC and fitted cross-sections. These differences are negligble for the plots of the poloidal cross sections of the quasisymmetric stellators shown in Fig. \ref{fig:poloidal plots}.

While there are very few visual discrepancies between the VMEC surfaces and the near-axis fits, we again note that near the axis, only $\psi_2$ is needed to accurately describe the surface \citep{Jorge2020Near-axisAxis}. Since $\psi_3$ is in Eq. \eqref{final well}, slight visual discrepancies in the third order fit may lead to larger discrepancies in the magnetic well. This motivates the need for a more direct way of calculating higher order surface coefficients.

In Fig. \ref{fig:magnetic_well_largerR} below, we plot the quantity $V_N'' = V''/V_0''$ as a function of the normalized toroidal flux, where $V''$ is the magnetic well calculated by VMEC, and $V_0''$ is the magnetic well calculated from Eq. \eqref{final well}. The W7-X equilibrium is the same, but for the two quasisymmetric stellarators, we choose boundary surfaces further from the axis, with $\psi_b \sim 0.003$ T m\textsuperscript{2}. For the quasisymmetric stellarators, even further from the axis ($\psi_N = 0.5$), the near-axis magnetic well remains approximately within 20\% of the VMEC magnetic well. We note that these surfaces represent a minor radius of about 0.1 m, and that similar discrepancies are found in Figure 1 of \citet{Landreman2020MagneticAxis}. The discrepancies are larger for the W7-X equilibrium because its boundary surface is significantly further from the axis. Finally, the first several VMEC surfaces for each design were excluded because the magnetic well calculated by VMEC changed very rapidly near the axis and so are more inaccurate.
\begin{figure}
    \centering
    \includegraphics[scale=0.75]{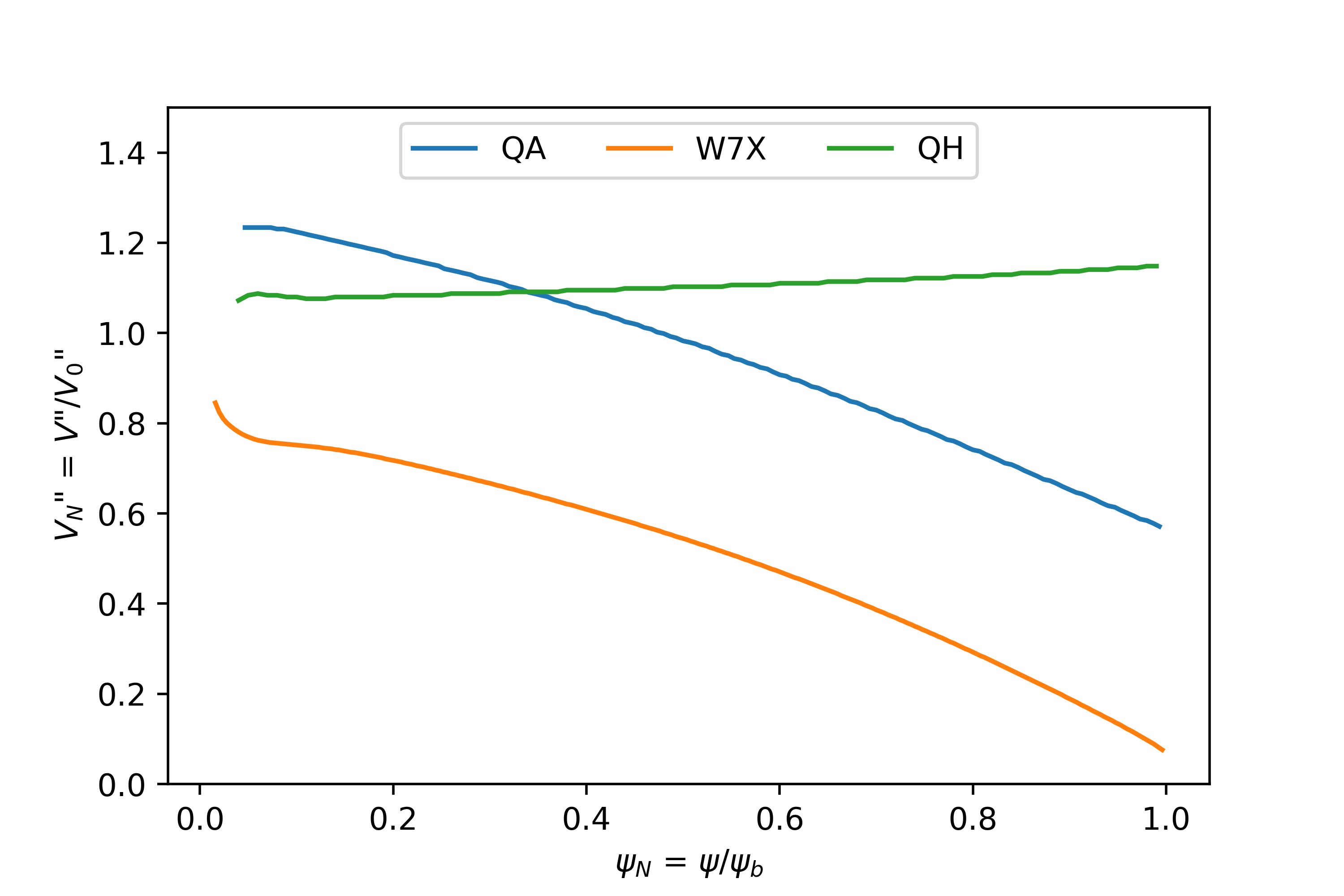}
    \caption{The ratio of the VMEC and near-axis magnetic well $V_N'' = V''/V_0''$ plotted over the normalized toroidal flux. The W7-X equilibrium is the same as in Fig. \ref{fig:magnetic_well}, but the quasisymmetric equilibriums have boundary surfaces further from the axis.}
    \label{fig:magnetic_well_largerR}
\end{figure}

\begin{figure}
    \centering
    \includegraphics[scale=0.7]{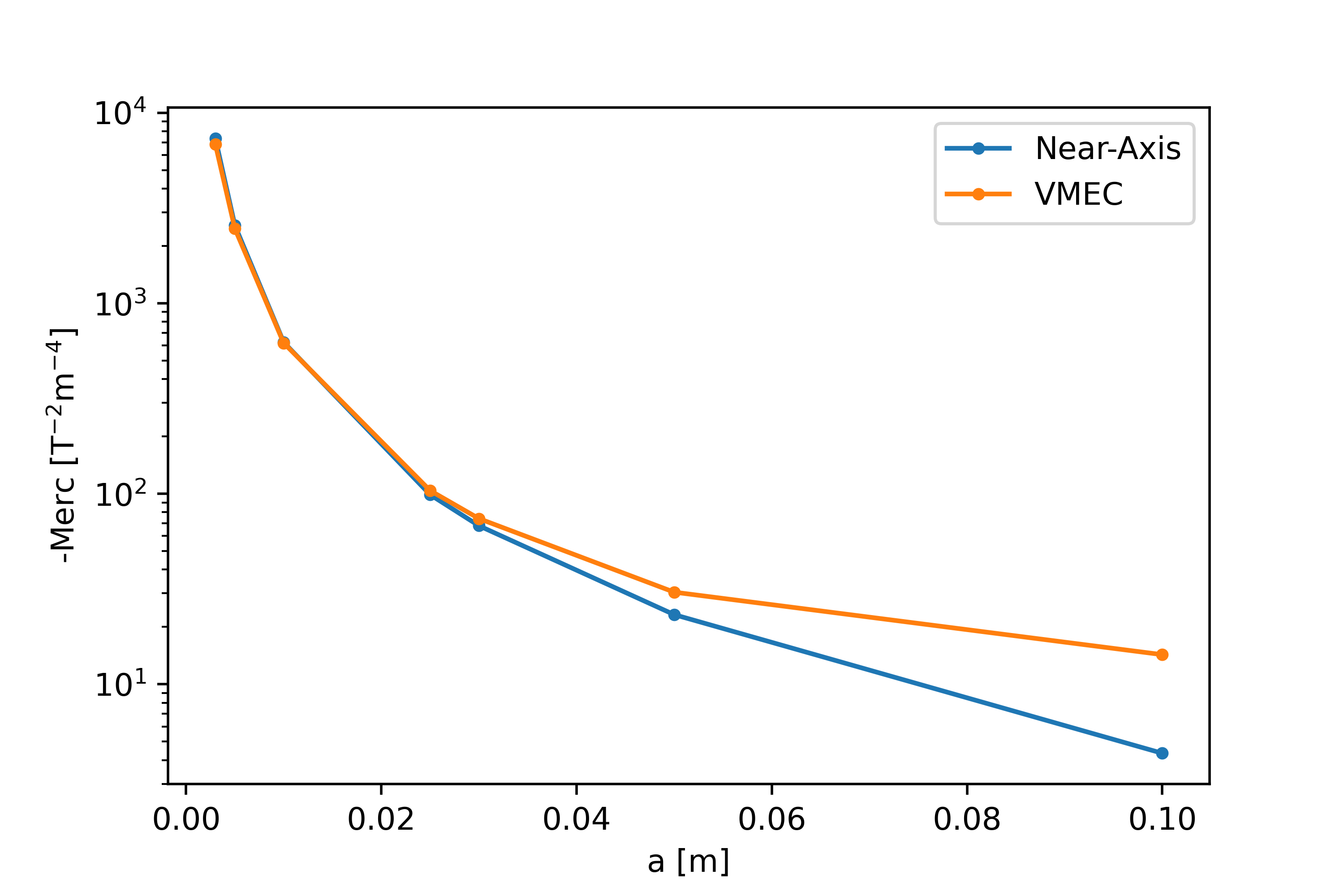}
    \caption{The lowest order criterion calculated from the near-axis expansion and the value of $D_{Merc}$ from VMEC. Each point represents a configuration evaluated at the boundary at different aspect ratios.}
    \label{fig:merceri crit larger r}
\end{figure}

Finally, after fitting the plasma parameters describing the quasi-axisymmetric stellarator equilibrium from Fig. \ref{fig:magnetic_well} for several of its surfaces, we numerically solve the system of equations in Eqs. \eqref{G1c} and \eqref{G1s}, and then evaluate the integral in Eq. \eqref{xi squared integral}. We then calculate the magnetic well and Eq. \eqref{Bsquared integral} and sum all the terms to calculate Mercier's criterion to lowest order. In Fig. \ref{fig:merceri crit larger r}, we compare our results for the lowest order criterion to the value of $D_{Merc}$ from VMEC. We can see that at smaller minor radii, the two results are very similar, but they diverge at larger radii. We note that at the largest radii $a = 0.1$ m, the expansion parameter $\epsilon \sim 0.18$.
\section{Conclusion}
In this work, we derive an expression for the on-axis magnetic well and lowest order Mercier's criterion using a direct coordinate near-axis approach. Although this derivation is inspired by the work originally presented in \citet{Mercier1964EquilibriumAxis}, we are able to simplify its calculations to a few essential steps, such as the cancelation of higher order terms using a particular form of $\mathbf{J} \times \mathbf{B} - \nabla p = 0$ . Furthermore, we compute the on-axis magnetic well using the SENAC code \citep{SENAC}, and verify it against several stellarator designs obtained using the code VMEC, including W7-X. We also numerically verify our expression for the lowest order criterion for the quasi-axisymmetric stellarator design from \citet{Landreman2019ConstructingOrder}. Additionally, we compare the expressions for the magnetic well in direct and inverse coordinate approach in the case of a tokamak. Since it has been shown that the direct coordinate approach is equivalent to the inverse approach to lowest order \citep{Jorge2020Near-axisAxis}, both formulations can be used to probe the stability of a variety of stellarator designs. In particular, we note that while we have used SENAC to fit VMEC equilibria for verification purposes, it is possible to use SENAC to generate equilibria near the axis and then calculate the magnetic well using the input parameters. Using this approach, a search of the parameter space for stellarator designs with optimized confinement properties, including for non-quasisymmetric geometries, is left for future work. However, we remark that higher order quantities (such as $\psi_3$) made substantial contributions to the magnetic well. As the values for $\psi_3$ depend on lower order quantities, an efficient search of the parameter space would require directly calculating $\psi_3$ from lower order quantities rather than from a fit. Therefore, future work will address more direct methods to calculate this quantity without the need for fitting large parameter spaces.

\section{Acknowledgements}
We wish to thank W. Sengupta and M. Landreman for many fruitful discussions about the contents of this manuscript. This work was supported by a grant from the United States Department of Energy (DEFG0293ER54197) and the Simons Foundation (560651, ML).

\appendix

\section{Mercier's Criterion Near the Magnetic Axis}\label{appB}
We now show that the lowest order terms in Mercier's criterion are those containing the magnetic well and the integral of $\Xi^2$, i.e., it reduces to only the terms in the second bracket of Eq. \eqref{Full criterion}. In the first brackets of Mercier's criterion in Eq. \eqref{Full criterion}, the shear term $d\iota/d\psi$ is of zeroth order in $\rho$. The $\mathbf{B} \cdot \mathbf{\Xi}$ term is also zeroth order in $\rho$. This is evident from our expression for $\Xi_{s}$ in Eq. \eqref{Xi s}. Since, to lowest order, $\mathbf{B}$ is aligned with the axis, $\mathbf{B} \cdot \mathbf{\Xi}$ can be written as
\begin{equation}
    \mathbf{B} \cdot \mathbf{\Xi} = 2\pi B_0^{5/2}e^{-\eta/2}\left(e^{\eta} G_{1s} \cos u-G_{1c} \sin u\right) + O(\epsilon).
\end{equation}
As $dS/|\nabla \psi|^3 \sim d\theta ds /[(\psi_2)^2\rho]$ is proportional to $1/\psi_2^2$, the resulting integrals of $\cos u/\psi_2^2$ and $\sin u/\psi_2^2$ over $\theta$ vanish over a full period. Since $\Xi_1$ is of order $\rho$ and $dS/|\nabla \psi|^3$ is of order $1/\rho^2$, the $\mathbf{B} \cdot \mathbf{\Xi}$ term is 0 at order $1/\rho$.

Using a similar argument, the final two terms in Eq. \eqref{Full criterion} containing $V''$ and $\Xi^2$ are of order $1/\rho^2$, and so are the lowest order terms. As stated after Eq. \eqref{xi squared integral}, the integral of $\Xi^2$ is proportional to $p_2^2$, while the magnetic well term is only proportional to $p_2$ in Eq. \eqref{Full criterion}. Therefore, in the limit of vanishing pressure gradient, the $\Xi^2$ integral will vanish faster, and the magnetic well becomes the dominant term.

\section{Solving Integrals with Cauchy's Residue Theorem}\label{appA}
In order to integrate Eq. \eqref{magnetic well} over $\theta$, we evaluate the integrals $\int_0^{2\pi} \cos 4u/(1+\mu\cos 2u)^3 du$, $\int_0^{2\pi} \cos 2u/(1+\mu\cos 2u)^3 du$, and $\int_0^{2\pi} 1/(1+\mu\cos 2u)^3 du$ using Cauchy's residue theorem, which states that if a function $f(z)$ is analytic, then its integral over a contour $\gamma$ is \citep{Ahlfors1979ComplexAnalysis}
\begin{equation}
    \int_{\gamma} f(z) dz = 2\pi i\sum_{a\in A}Res\left[f(a)\right],
\end{equation}%
where $A$ is the set of all the poles of $f(z)$ enclosed by $\gamma$.
To solve the integral of $\int_0^{2\pi} \cos 4u/(1+\mu\cos 2u)^3 du$, we first convert all the cosine functions into their complex exponential forms. We then define the complex valued parameter $z=e^{2iu}$. An integral over a full period of $u$ is equivalent to a contour integral over the unit circle defined by $e^{2iu}$. A change of variables results in
\begin{equation}\label{complex_integral_1}
   \int_0^{2\pi} \frac{\cos 4u}{(1+\mu\cos 2u)^3} du = -2i \oint_\gamma dz \frac{z^4+1}{(\mu z^2 + 2z + \mu)^4},
\end{equation}%
where $\gamma$ is the contour defined by $e^{2iu}$. The denominator of Eq. \eqref{complex_integral_1} is quadratic with roots $z_1 = (-1+\sqrt{1-\mu^2})/\mu$ and $z_2 = (-1-\sqrt{1-\mu^2})/\mu$. We note that $z_1z_2 = 1$. As a result, $z_1$ and $z_2$ cannot both be outside or inside the unit circle. Since $z_1 < z_2$, only $z_1$ is enclosed by the contour. Thus, we only consider the residue at $z = z_1$. By the residue theorem, the value of the contour integral of Eq. \eqref{complex_integral_1} is then
\begin{equation}
    2\pi i \ Res\left(-2i\frac{z^4+1}{(\mu z^2 + 2z + \mu)^4}\right)\bigg\vert_{z=z_1} = \sum_\gamma \frac{3\pi\mu^2}{2(1-\mu^2)^{5/2}},
\end{equation}%
where the summation is due to the complex parameter $z$ going twice around the pole.
We take a similar approach to evaluate the other two integrals, yielding
\begin{equation}
    \int_0^{2\pi} \frac{\cos 2u}{(1+\mu\cos 2u)^3} du = -\sum_\gamma \frac{3\pi\mu}{2(1-\mu^2)^{5/2}},
\end{equation}
and
\begin{equation}
    \int_0^{2\pi} \frac{1}{(1+\mu\cos 2u)^3} du = \sum_\gamma \frac{\pi(2+\mu^2)}{2(1-\mu^2)^{5/2}}.
\end{equation}

\bibliographystyle{jpp}

\bibliography{references}

\end{document}